\documentclass[aip,
reprint,
superscriptaddress]{revtex4-1}

\usepackage{graphicx}
\usepackage{amsmath}
\usepackage[colorlinks=true,urlcolor=blue,citecolor=black]{hyperref}
\usepackage[]{natbib}
\usepackage{xcolor}

\begin{document}

\title{Protective capping of topological surface states of intrinsically insulating Bi$_2$Te$_3$}

\author{Katharina Hoefer}
\email{katharina.hoefer@cpfs.mpg.de}
\affiliation{Max Planck Institute for
Chemical Physics of Solids, N\"{o}thnitzer Strasse 40, Dresden 01187, Germany}
\author{Christoph Becker}\affiliation{Max Planck Institute for
Chemical Physics of Solids, N\"{o}thnitzer Strasse 40, Dresden 01187, Germany}
\author{Steffen Wirth}\affiliation{Max Planck Institute for
Chemical Physics of Solids, N\"{o}thnitzer Strasse 40, Dresden 01187, Germany}
\author{Liu Hao Tjeng}
\email{hao.tjeng@cpfs.mpg.de}
\affiliation{Max Planck Institute for
Chemical Physics of Solids, N\"{o}thnitzer Strasse 40, Dresden 01187, Germany}

\date{\today}


\begin{abstract}

We have  identified epitaxially grown elemental Te as a capping material that is suited to protect the topological
surface states of intrinsically insulating Bi$_2$Te$_3$. By using angle-resolved
photoemission, we were able to show  that the Te overlayer leaves the dispersive bands of the surface states intact and that it does
not alter the chemical potential of the Bi$_2$Te$_3$ thin film. From \textit{in-situ} four-point contact measurements, we observed that the conductivity of the capped film is still mainly determined by the metallic surface
states and that the contribution of the capping layer is minor. Moreover, the Te overlayer can be annealed away in vacuum to produce a clean Bi$_2$Te$_3$ surface in its pristine state even after the exposure of the capped film to air. Our findings will facilitate well-defined and reliable \textit{ex-situ} experiments on the
properties of  Bi$_2$Te$_3$ surface states with nontrivial topology.

\end{abstract}

\maketitle

Topological insulators (TI) form a novel state of matter. They have an electrically insulating bulk, simultaneously, they have a necessarily metallic surface. The surface states have massless Dirac dispersions, and the charge carriers are
protected from back-scattering by time reversal symmetry\cite{Kane2005,Fu2007,Zhang2009a}.
These properties make TI promising for advanced spintronic applications as well as for the realization of novel quantum particles\cite{Fu2008}.

At present, it is very difficult to experimentally measure the topological surface states without being disturbed by
signals from the bulk or surface impurities in materials such as Bi$_2$Se$_3$ and Bi$_2$Te$_3$.
This is mainly because the Fermi surface enclosed by these surface states forms only a small fraction of the surface
Brillouin zone, therefore,  the number of charge carriers with interesting topological properties is only of the
order of a few 10$^{12}$\,cm$^{-2}$. This, in turn, sets a very strict constraint in that the concentration of impurities
at the surface must be much less than 1\%, furthermore, the concentration of defects in
the bulk has to be well below the ppm level depending on the thickness of the material.

Recently, we have been able to prepare Bi$_2$Te$_3$ films, using molecular beam epitaxy, whose  conductivity  is
dominated by the surface\cite{Hoefer2014}. Growth, structural characterization using  reflection high energy
electron diffraction (RHEED) and low energy electron diffraction (LEED), electronic structure determination by angle-resolved and X-ray photoelectron spectroscopy (ARPES and XPS, respectively), and resistivity measurements
were all performed  \textit{in-situ} under ultrahigh vacuum (UHV) conditions.  We found that the chemical
potential straddles through the surface states and lies inside the bulk band gap, implying that the surface is metallic and the bulk is a good insulator.

Separately, we also discovered that exposure to tiny amounts of air (water) causes the bulk conduction band to get filled
with electrons. As a result, the conductivity increases and is no longer determined by the surface
states alone\cite{Hoefer2014}. Therefore, to facilitate \textit{ex-situ} experiments and fabricate devices
using the nontrivial properties of  surface states,  methods need to be developed to cap and protect these
surface states \cite{Kong2011,Ngabonziza2015,Hoefer2014}. Some studies have attempted to use amorphous
Se or Te as a capping material \cite{Harrison2014,Virwani2014}. However,  it remains  unclear whether  the intrinsic properties of the surface states have been affected. One thing that was clear was that the stoichiometry of the
films has been altered after the removal of the capping material\cite{Harrison2014,Virwani2014}.

In this study, we report on the use of epitaxially grown Te as a capping material. Our strategy is to deposit the material
under UHV conditions in a layer-by-layer fashion and monitor the properties of the Bi$_2$Te$_3$ film, especially those of the surface states using \textit{in-situ} XPS, ARPES, and conductivity measurements. We also tested  whether the Te capping can be removed  to restore the Bi$_2$Te$_3$ film with the surface states in their
pristine condition and whether this can also be achieved after the exposure of the capped film to air.

Bi$_2$Te$_3$ thin films of typical 20\,QL (quintuple layer; 1\,QL $\sim$1\,nm)  were grown using molecular beam epitaxy (MBE) under UHV conditions (low 10$^{-10}$\,mbar) on insulating epi-polished and vacuum-annealed BaF$_2 (1\,1\,1)$ substrates. The details of the growth conditions for the Bi$_2$Te$_3$ films and the experimental setup can be found elsewhere\cite{Hoefer2014}. Te capping was conducted in steps, resulting in a Te overlayer thickness of 1\,u.c.(unit cell, 6\,{\AA}),  2\,u.c (12\,{\AA}),
5\,u.c. (30\,{\AA}),  10\,u.c. (60\,{\AA}), and 20\,u.c. (120\,{\AA}). The Te flux rate was set to 1\,{\AA}/min
with the Bi$_2$Te$_3$ film kept at room temperature.

\begin{figure*}[htb]
 \centerline{
  \includegraphics[width=1\textwidth]{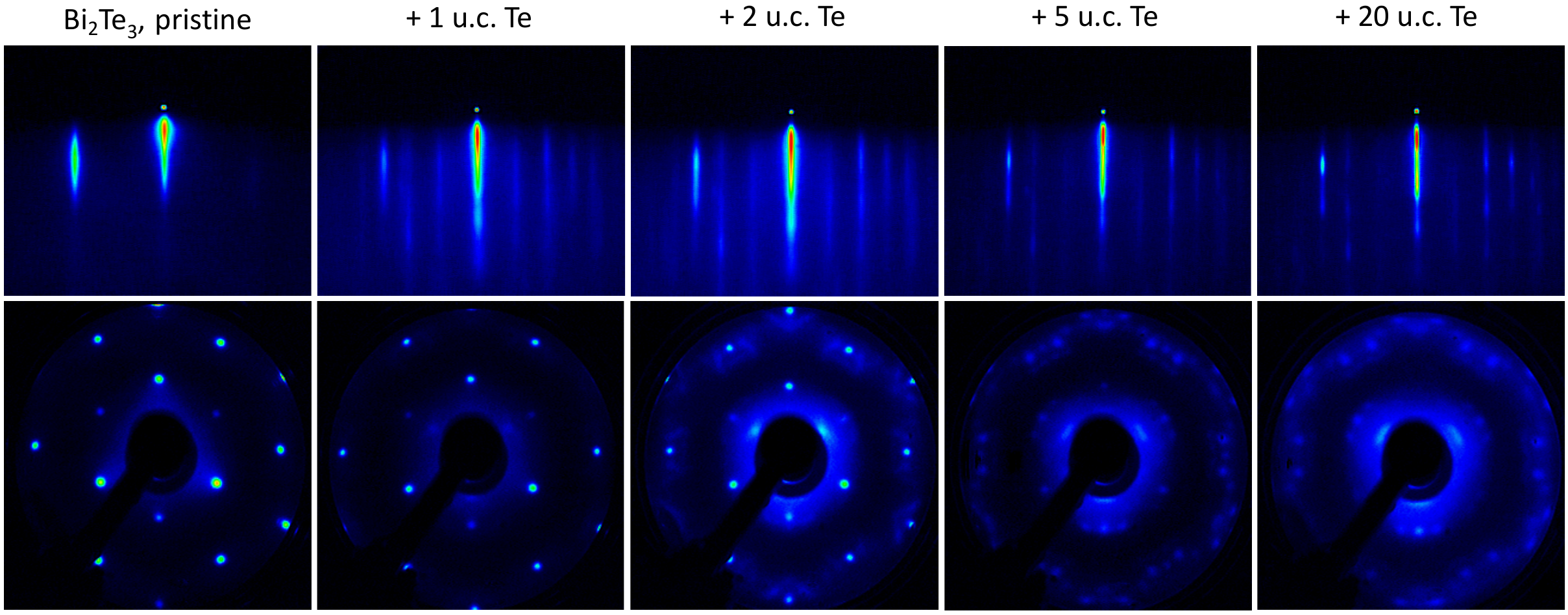}}
  \caption{RHEED and LEED patterns of a pristine 20\,QL Bi$_2$Te$_3$ film and the same film at different stages
(1\,u.c. $\sim$6\,{\AA}, 2\,u.c., 5\,u.c. and 20\,u.c.) of  Te capping. The electron energy is set at 17.5\,keV for RHEED and 52\,eV for LEED measurements.}\label{RHEED_LEED}
\end{figure*}

The RHEED and LEED patterns of a pristine 20\,QL Bi$_2$Te$_3$ film for different thicknesses of Te capping are
shown in Fig.~\ref{RHEED_LEED}. The LEED of the noncapped sample shows three-fold symmetry as expected for single-domain Bi$_2$Te$_3$. Upon capping with 1 and 2\,u.c. Te, additional RHEED streaks and LEED spots appear. The features are sharp, and with  the three-fold symmetry still clearly visible in the LEED, we can conclude
that  the Te overlayer can be epitaxially grown on the  Bi$_2$Te$_3$ film. For thicker Te films, e.g. 5 and 20\,u.c., multiple LEED patterns appear, and the features become less intense compared to the
background. Nonetheless, they remain sharp, as seen in the LEED photographs taken with 26\,eV electrons as shown in  Fig.~S1 of the supplemental material. These observations indicate an epitaxial but multidomain growth mode for thicker
Te overlayers, this is probably related to the 1.6\% lattice mismatch between  Te and Bi$_2$Te$_3$ and the need to
relax the strain in the Te overlayer.

\begin{figure}[htb]
  \centerline{
  \includegraphics[width=1\columnwidth]{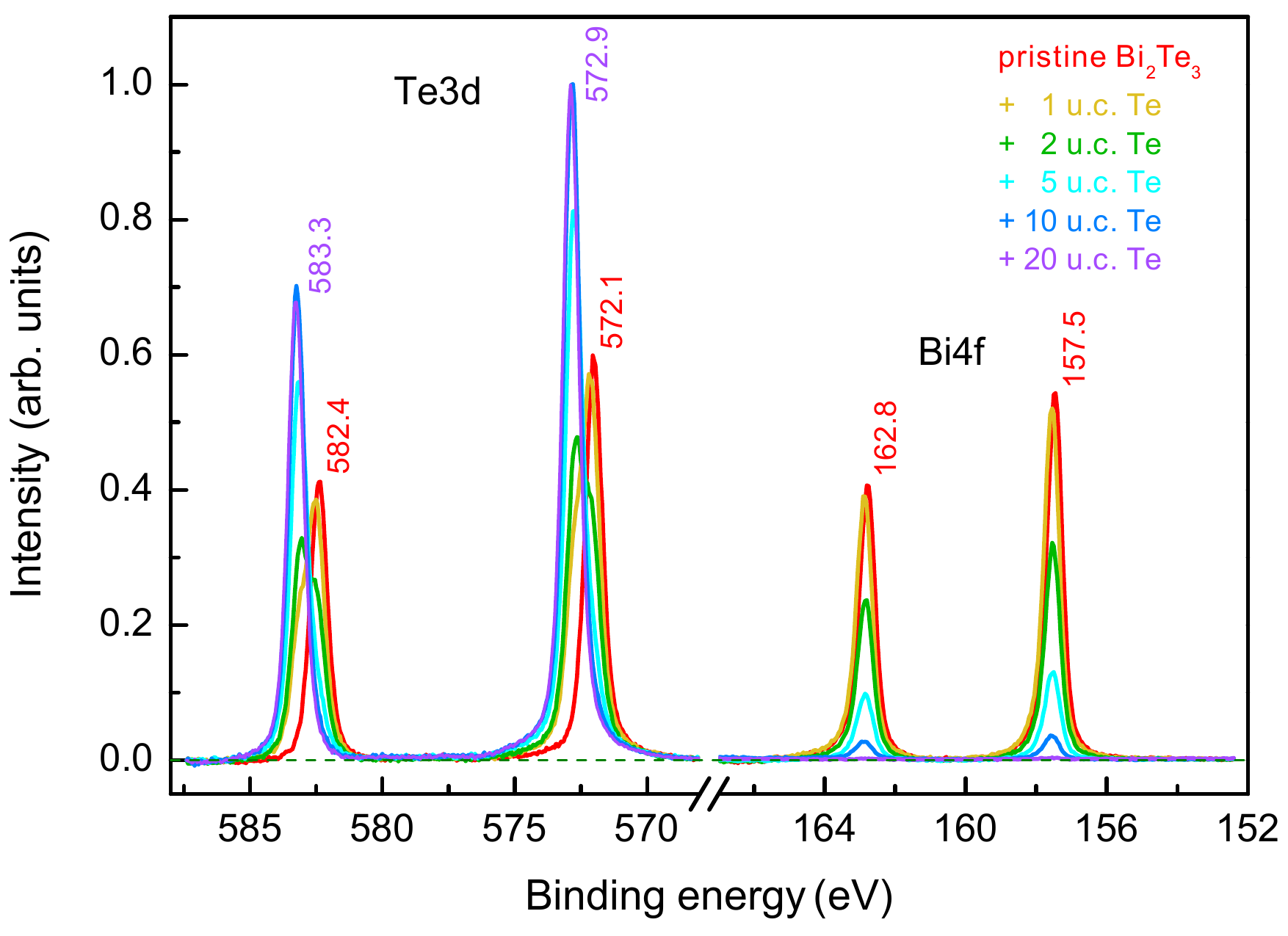}}
  \caption{X-ray photoelectron spectroscopy of the Te\,3d and Bi\,4f core levels of the Bi$_2$Te$_3$ sample with Te cappings of different thicknesses.}
  \label{XPS}
\end{figure}

XPS measurements reveal very narrow and symmetric Te\,3d and Bi\,4f core level lines for the pristine Bi$_2$Te$_3$ film, as indicated by the  red curve in Fig.~\ref{XPS}. With increasing Te thickness, the intensity of the Bi\,4f core level decreases
without any changes in the line shape or energy position. This decrease is exponential with the 1/e value
of 22\,{\AA} ($\sim$3.8\,u.c), which well fits the typical mean free paths of 1-1.5\,keV photoelectrons\cite{Huefner2003}, indicating that the Te capping forms rather flat overlayers without too many pinholes.
The Bi of  Bi$_2$Te$_3$ is also chemically not affected by the Te capping. The Te\,3d line develops a shoulder at around 0.8\,eV higher binding energies upon the deposition of the Te overlayer. This shoulder can be attributed to
elemental Te\cite{Moulder1992}. The 12\,nm thick (20\, u.c.)  Te capping fully suppresses the  Bi$_2$Te$_3$ signal; only the elemental Te\,3d$_{3/2}$ and Te\,3d$_{5/2}$ peaks are visible at $\sim$583\,eV and $\sim$573\,eV, respectively.

\begin{figure*}[htb]
  \centerline{
  \includegraphics[width=0.9\textwidth]{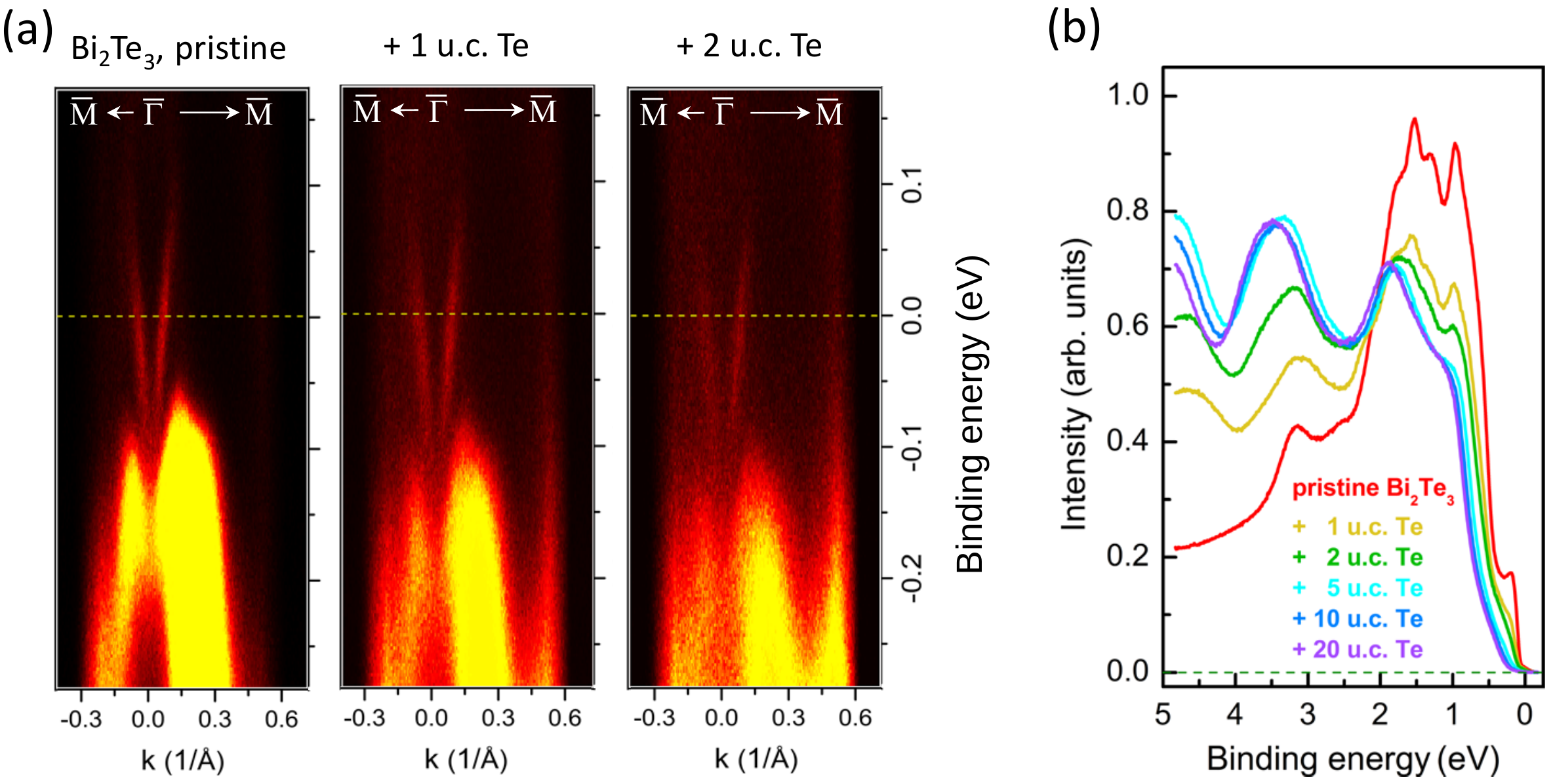}}
  \caption{(a) ARPES spectra of a 20\,QL Bi$_2$Te$_3$ film and the same film capped with 6\,{\AA} (1\,u.c.) and
              12\,{\AA} (2\,u.c.), respectively. The data were taken along the $\bar{\Gamma}-\bar{M}$ direction
              at room temperature. (b) Angle integrated UPS spectra of the pristine Bi$_2$Te$_3$ and for different
              Te capping layers.}\label{ARPES}
\end{figure*}

Figure~\ref{ARPES}\,(a) shows the ARPES spectra using He\,I light (21.2\,eV) taken along the
$\bar{\Gamma}-\bar{M}$ direction of the pristine 20\,QL thick Bi$_2$Te$_3$ film at room temperature and the
corresponding spectra with capping layers of 6\,{\AA} (1\,u.c.) and 12\,{\AA} (2\,u.c.). One can clearly observe
the Dirac cone and the topological surface states for the three measurements. Most importantly, for all  steps of Te deposition, the Fermi level (zero binding energy) intersects with the surface states only and not with the bulk valence band
nor with the bulk conduction band of the Bi$_2$Te$_3$ film. This indicates that surface states of the pristine film
are not affected by the Te capping, i.e., not only are the dispersive bands of the topologically nontrivial surface
states intact but also their filling remains the same. In particular, the latter is very remarkable in light of the fact that
the amount of charge carriers with  interesting topological properties is only of the order of a few 10$^{12}$\,cm$^{-2}$, i.e., of the order of 0.01 per surface unit cell. Thus, our Te capping method does not cause doping of the Bi$_2$Te$_3$ surface. This observation is in agreement with a very recent work on crystalline Te capping of (Bi,Sb)$_2$Te$_3$ thin films\cite{Park2015}.

For Te capping thicker than $\sim$12\,{\AA} (2\,u.c.), we can no longer see the signal from Bi$_2$Te$_3$. This
can be attributed to the very small probing depth, of the order of 5\,{\AA}, of the ARPES techniques using He\,I (21.2\,eV)\cite{Tjeng1997}. Figure~\ref{ARPES}\,(b) shows
the angle-integrated UPS (ultraviolet photoelectron spectroscopy) spectrum of  pristine Bi$_2$Te$_3$ with different  Te capping layers. It is evident that
the Bi$_2$Te$_3$ states near the Fermi level are no longer visible for a capping thicker than 30\,{\AA} (5\,u.c.)
and that the spectra for 30\,{\AA} (5\,u.c.), 60\,{\AA} (10\,u.c.), and 120\,{\AA} (20\,u.c.) are similar and essentially
given by elemental Te.

\begin{figure}[htb]
  \centerline{
  \includegraphics[width=1\columnwidth]{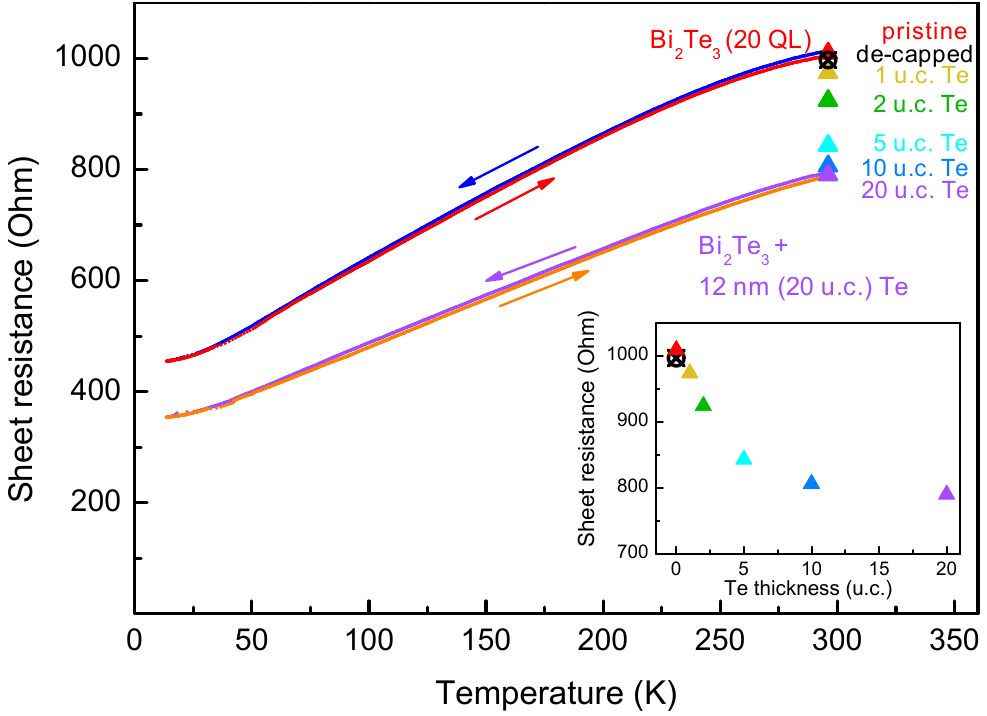}}
  \caption{\textit{In-situ} sheet resistance measurements of the pristine and capped 20\,QL thin Bi$_2$Te$_3$ film.
The temperature dependent cool-down and warm-up curves are shown for the pristine Bi$_2$Te$_3$ (blue and red)
and the film capped with 12\,nm Te (violet and orange), respectively. The triangles indicate the room temperature sheet resistance values for Te cappings of different  thicknesses. The black crossed circle indicated the room temperature
value of the sheet resistance after the removal of the capping layer by thermal desorption at 220$^{\circ}$C. The inset displays the sheet resistance versus the Te capping layer thickness at room temperature.}
  \label{R-T}
\end{figure}

Subsequently, the influence of the Te capping on the electrical transport  properties of the Bi$_2$Te$_3$ film were investigated.
We conducted the experiments \textit{in-situ}, i.e., the contacting and temperature-dependent resistivity
measurements were all performed under UHV conditions, thereby excluding any external effects that might occur owing to degradation when the film would have been exposed to air. The results are presented in Figure~\ref{R-T}. The temperature-dependence of the sheet resistance was investigated for the pristine Bi$_2$Te$_3$ film (blue and red curves)
and for the film fully capped with 12\,nm Te (violet and orange curves). For the intermediate Te thicknesses
only the sheet resistance values at room temperature are shown (triangles). The pristine film shows  metallic behavior over the entire temperature range without any sign of degradation of the contacts during the temperature
cycle. Upon applying of the Te capping, the resistivity decreases gradually but not dramatically. Most of the
conductivity is still determined by the surface states of the pristine Bi$_2$Te$_3$ film.

It should be noted that the extra conductivity owing to the Te capping cannot be accounted for in terms of a simple parallel conductivity due a thin slab of Te with Te bulk properties. The sheet resistance drops, for example, by about 200\,Ohm with a 6\,nm capping, which is much more than the few Ohm reduction expected on the basis of the specific resistivity of bulk Te ($\rho_{\perp}$ = 1.5\,mOhmm perpendicular to the c-axis at 20$^{\circ}$C). We speculate that the strain or polarizability exerted \cite{Hesper1997} by the adjacent Bi$_2$Te$_3$ influences the properties of the Te layers closest to the interface in such a manner that those Te layers becomes much more conducting  than bulk Te. We note on the other hand that the sheet resistance drop between 6\,nm and 12\,nm capping is only a few Ohm, well within the expectations for an additional 6\,nm thin Te slab having the Te bulk resistivity. We therefore can infer that the Te film is essentially bulk like, except for the Te layers within a few nm from the interface with the Bi$_2$Te$_3$. Since the photoemission data showed no shift of the Fermi level due to the Te capping, we can assume that the resistance of the Bi$_2$Te$_3$ surface states remains to have the $\approx$1\,kOhm value and that the resistance drop is caused by the parallel shunt by the first 6\,nm Te having about $\approx$4\,kOhm resistance (with any additional Te layers having the bulk Te properties). In any case, we can be assured that the sheet resistance of the Bi$_2$Te$_3$ with capping is still overwhelmingly given by that of the intrinsic surface states.

\begin{figure}[t]
  \centerline{
  \includegraphics[width=1\columnwidth]{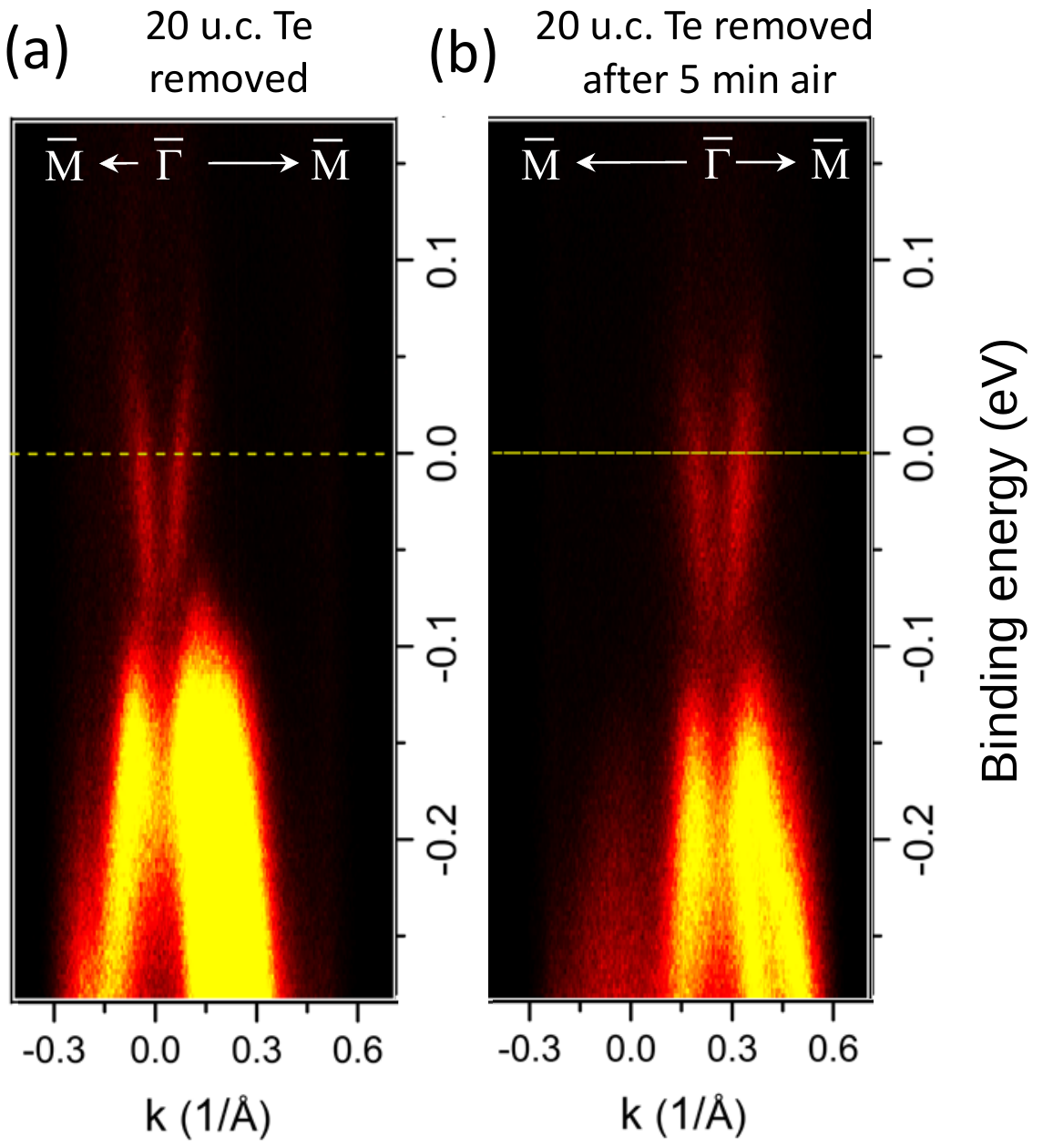}}
  \caption{ARPES spectra of two different Bi$_2$Te$_3$ films after removal of a 12\,nm Te capping.
              (a) Te capped sample kept in UHV.
              (b) Te capped sample  exposed to air for 5\,min.}
  \label{removal}
\end{figure}

The observation that the Te capping does not affect the surface states of the Bi$_2$Te$_3$ film  also means that
there is no chemical reaction between the Te capping and  Bi$_2$Te$_3$. This suggests
that it should be possible to remove the capping layer by physical means so as to recover the pristine state of the
Bi$_2$Te$_3$ surface. We have tested this possibility by gradually heating up the capped film while simultaneously monitoring the
RHEED pattern. The substrate temperature was first ramped to 180$^{\circ}$C and then
increased in steps of 10\,K with a waiting time of 5\,min at each step. At 220$^{\circ}$C, the additional RHEED
streaks, originating from  Te, start to vanish, and after 10\,min  the original Bi$_2$Te$_3$ RHEED pattern was
recovered (cf. Fig.~S2). Fig~\ref{removal}\,(a) shows the ARPES spectrum of this annealed thin film. All features of the
pristine Bi$_2$Te$_3$ film are fully restored. The high contrast of the band structure features relative to the background and the similar position of the Fermi level crossings indicate that there are no visible changes in the surface morphology and stoichiometry. It should also be noted that the removal of the Te capping is complete,
as evidenced by the XPS measurements shown in Fig.~S3 of the supplemental material. The room-temperature sheet resistance of this annealed film (black crossed circle in Fig.~\ref{R-T}) is identical to that of the film prior to
capping. This serves as further proof that  capping and  de-capping can be conducted such that the properties of the pristine topologically nontrivial surface states are preserved.
In contrast to previous studies\cite{Harrison2014,Virwani2014}, where problems with the stoichiometry were reported, we did not observe changes in the stoichiometry of the Bi$_2$Te$_3$ surface after capping and de-capping Te (see Fig.~\ref{removal} and Fig.~S3).

The issue is now to show that  capping does protect  Bi$_2$Te$_3$ against degradation upon exposure to
ambient conditions. Toward this end, we exposed a capped film to air for at least 5\,min at room temperature. After inserting the film
back into the UHV system, we conducted the above-described annealing process to remove the Te capping.  Figure~\ref{removal}\,(b) shows the result, where one can clearly observe that the ARPES spectrum is essentially the same as that of the pristine Bi$_2$Te$_3$ film shown in Fig.~\ref{ARPES}\,(a). This result demonstrates that the capping is leakproof, i.e., the number of pinholes or cracks in the Te overlayer are apparently negligible. It should be noted that our findings are also facilitated by the fact that pure Te thin films are rather inert against oxidation. As has been reported earlier, exposure to air for longer duration does not necessarily result in the formation of TeO$_2$ \cite{Musket1978}. This also means that if oxidation  is occurring, it will be a very slow process \cite{Lee1983} and will therefore  not affect the
properties of the deeply covered Bi$_2$Te$_3$ topological surface states during even prolonged \textit{ex-situ}
experiments.

In conclusion, we  showed that epitaxially grown elemental Te is an ideal capping material to protect the topological
surface states of intrinsically insulating Bi$_2$Te$_3$.  Our angle-resolved photoemission experiments established
that the Te capping leaves the band dispersions and  Fermi level position of the surface states intact. Our \textit{in-situ} four point contact measurements verified that the electrical properties of the surface states can still be clearly and directly detected despite the presence of the capping. We have demonstrated that the Te
capping does indeed protect the Bi$_2$Te$_3$ surface against the influence of ambient conditions. Moreover, we have shown that the capping can be easily removed by thermal annealing in vacuum, thereby restoring the Bi$_2$Te$_3$ surface to its pristine state. These results will therefore enable a wide range of well-defined
and reliable \textit{ex-situ} and surface/interface sensitive experiments using the Bi$_2$Te$_3$ surface states with
nontrivial topology.

\bibliography{references_Te}

\begin{thebibliography}{16}%
\makeatletter
\providecommand \@ifxundefined [1]{%
 \@ifx{#1\undefined}
}%
\providecommand \@ifnum [1]{%
 \ifnum #1\expandafter \@firstoftwo
 \else \expandafter \@secondoftwo
 \fi
}%
\providecommand \@ifx [1]{%
 \ifx #1\expandafter \@firstoftwo
 \else \expandafter \@secondoftwo
 \fi
}%
\providecommand \natexlab [1]{#1}%
\providecommand \enquote  [1]{``#1''}%
\providecommand \bibnamefont  [1]{#1}%
\providecommand \bibfnamefont [1]{#1}%
\providecommand \citenamefont [1]{#1}%
\providecommand \href@noop [0]{\@secondoftwo}%
\providecommand \href [0]{\begingroup \@sanitize@url \@href}%
\providecommand \@href[1]{\@@startlink{#1}\@@href}%
\providecommand \@@href[1]{\endgroup#1\@@endlink}%
\providecommand \@sanitize@url [0]{\catcode `\\12\catcode `\$12\catcode
  `\&12\catcode `\#12\catcode `\^12\catcode `\_12\catcode `\%12\relax}%
\providecommand \@@startlink[1]{}%
\providecommand \@@endlink[0]{}%
\providecommand \url  [0]{\begingroup\@sanitize@url \@url }%
\providecommand \@url [1]{\endgroup\@href {#1}{\urlprefix }}%
\providecommand \urlprefix  [0]{URL }%
\providecommand \Eprint [0]{\href }%
\providecommand \doibase [0]{http://dx.doi.org/}%
\providecommand \selectlanguage [0]{\@gobble}%
\providecommand \bibinfo  [0]{\@secondoftwo}%
\providecommand \bibfield  [0]{\@secondoftwo}%
\providecommand \translation [1]{[#1]}%
\providecommand \BibitemOpen [0]{}%
\providecommand \bibitemStop [0]{}%
\providecommand \bibitemNoStop [0]{.\EOS\space}%
\providecommand \EOS [0]{\spacefactor3000\relax}%
\providecommand \BibitemShut  [1]{\csname bibitem#1\endcsname}%
\let\auto@bib@innerbib\@empty
\bibitem [{\citenamefont {Kane}\ and\ \citenamefont {Mele}(2005)}]{Kane2005}%
  \BibitemOpen
  \bibfield  {author} {\bibinfo {author} {\bibfnamefont {C.~L.}\ \bibnamefont
  {Kane}}\ and\ \bibinfo {author} {\bibfnamefont {E.~J.}\ \bibnamefont
  {Mele}},\ }\bibfield  {title} {\enquote {\bibinfo {title} {Z$_{2}$
  topological order and the quantum spin hall effect},}\ }\href
  {http://link.aps.org/doi/10.1103/PhysRevLett.95.146802} {\bibfield  {journal}
  {\bibinfo  {journal} {Phys. Rev. Lett.}\ }\textbf {\bibinfo {volume} {95}},\
  \bibinfo {pages} {146802} (\bibinfo {year} {2005})}\BibitemShut {NoStop}%
\bibitem [{\citenamefont {Fu}, \citenamefont {Kane},\ and\ \citenamefont
  {Mele}(2007)}]{Fu2007}%
  \BibitemOpen
  \bibfield  {author} {\bibinfo {author} {\bibfnamefont {L.}~\bibnamefont
  {Fu}}, \bibinfo {author} {\bibfnamefont {C.~L.}\ \bibnamefont {Kane}}, \ and\
  \bibinfo {author} {\bibfnamefont {E.~J.}\ \bibnamefont {Mele}},\ }\bibfield
  {title} {\enquote {\bibinfo {title} {Topological insulators in three
  dimensions},}\ }\href {http://link.aps.org/doi/10.1103/PhysRevLett.98.106803}
  {\bibfield  {journal} {\bibinfo  {journal} {Phys. Rev. Lett.}\ }\textbf
  {\bibinfo {volume} {98}},\ \bibinfo {pages} {106803} (\bibinfo {year}
  {2007})}\BibitemShut {NoStop}%
\bibitem [{\citenamefont {Zhang}\ \emph {et~al.}(2009)\citenamefont {Zhang},
  \citenamefont {Liu}, \citenamefont {Qi}, \citenamefont {Dai}, \citenamefont
  {Fang},\ and\ \citenamefont {Zhang}}]{Zhang2009a}%
  \BibitemOpen
  \bibfield  {author} {\bibinfo {author} {\bibfnamefont {H.}~\bibnamefont
  {Zhang}}, \bibinfo {author} {\bibfnamefont {C.-X.}\ \bibnamefont {Liu}},
  \bibinfo {author} {\bibfnamefont {X.-L.}\ \bibnamefont {Qi}}, \bibinfo
  {author} {\bibfnamefont {X.}~\bibnamefont {Dai}}, \bibinfo {author}
  {\bibfnamefont {Z.}~\bibnamefont {Fang}}, \ and\ \bibinfo {author}
  {\bibfnamefont {S.-C.}\ \bibnamefont {Zhang}},\ }\bibfield  {title} {\enquote
  {\bibinfo {title} {Topological insulators in {Bi$_2$Se$_3$}, {Bi$_2$Te$_3$}
  and {Sb$_2$Te$_3$} with a single {Dirac} cone on the surface},}\ }\href
  {http://dx.doi.org/10.1038/nphys1270} {\bibfield  {journal} {\bibinfo
  {journal} {Nat Phys}\ }\textbf {\bibinfo {volume} {5}},\ \bibinfo {pages}
  {438--442} (\bibinfo {year} {2009})}\BibitemShut {NoStop}%
\bibitem [{\citenamefont {Fu}\ and\ \citenamefont {Kane}(2008)}]{Fu2008}%
  \BibitemOpen
  \bibfield  {author} {\bibinfo {author} {\bibfnamefont {L.}~\bibnamefont
  {Fu}}\ and\ \bibinfo {author} {\bibfnamefont {C.~L.}\ \bibnamefont {Kane}},\
  }\bibfield  {title} {\enquote {\bibinfo {title} {Superconducting proximity
  effect and {Majorana} fermions at the surface of a topological insulator},}\
  }\href {\doibase 10.1103/PhysRevLett.100.096407} {\bibfield  {journal}
  {\bibinfo  {journal} {Phys. Rev. Lett.}\ }\textbf {\bibinfo {volume} {100}},\
  \bibinfo {pages} {096407} (\bibinfo {year} {2008})}\BibitemShut {NoStop}%
\bibitem [{\citenamefont {Hoefer}\ \emph {et~al.}(2014)\citenamefont {Hoefer},
  \citenamefont {Becker}, \citenamefont {Rata}, \citenamefont {Swanson},
  \citenamefont {Thalmeier},\ and\ \citenamefont {Tjeng}}]{Hoefer2014}%
  \BibitemOpen
  \bibfield  {author} {\bibinfo {author} {\bibfnamefont {K.}~\bibnamefont
  {Hoefer}}, \bibinfo {author} {\bibfnamefont {C.}~\bibnamefont {Becker}},
  \bibinfo {author} {\bibfnamefont {D.}~\bibnamefont {Rata}}, \bibinfo {author}
  {\bibfnamefont {J.}~\bibnamefont {Swanson}}, \bibinfo {author} {\bibfnamefont
  {P.}~\bibnamefont {Thalmeier}}, \ and\ \bibinfo {author} {\bibfnamefont
  {L.~H.}\ \bibnamefont {Tjeng}},\ }\bibfield  {title} {\enquote {\bibinfo
  {title} {Intrinsic conduction through topological surface states of
  insulating {Bi$_2$Te$_3$} epitaxial thin films},}\ }\href@noop {} {\bibfield
  {journal} {\bibinfo  {journal} {Proceedings of the National Academy of
  Sciences}\ }\textbf {\bibinfo {volume} {111}},\ \bibinfo {pages}
  {14979--14984} (\bibinfo {year} {2014})}\BibitemShut {NoStop}%
\bibitem [{\citenamefont {Kong}\ \emph {et~al.}(2011)\citenamefont {Kong},
  \citenamefont {Cha}, \citenamefont {Lai}, \citenamefont {Peng}, \citenamefont
  {Analytis}, \citenamefont {Meister}, \citenamefont {Chen}, \citenamefont
  {Zhang}, \citenamefont {Fisher}, \citenamefont {Shen},\ and\ \citenamefont
  {Cui}}]{Kong2011}%
  \BibitemOpen
  \bibfield  {author} {\bibinfo {author} {\bibfnamefont {D.}~\bibnamefont
  {Kong}}, \bibinfo {author} {\bibfnamefont {J.~J.}\ \bibnamefont {Cha}},
  \bibinfo {author} {\bibfnamefont {K.}~\bibnamefont {Lai}}, \bibinfo {author}
  {\bibfnamefont {H.}~\bibnamefont {Peng}}, \bibinfo {author} {\bibfnamefont
  {J.~G.}\ \bibnamefont {Analytis}}, \bibinfo {author} {\bibfnamefont
  {S.}~\bibnamefont {Meister}}, \bibinfo {author} {\bibfnamefont
  {Y.}~\bibnamefont {Chen}}, \bibinfo {author} {\bibfnamefont {H.-J.}\
  \bibnamefont {Zhang}}, \bibinfo {author} {\bibfnamefont {I.~R.}\ \bibnamefont
  {Fisher}}, \bibinfo {author} {\bibfnamefont {Z.-X.}\ \bibnamefont {Shen}}, \
  and\ \bibinfo {author} {\bibfnamefont {Y.}~\bibnamefont {Cui}},\ }\bibfield
  {title} {\enquote {\bibinfo {title} {Rapid surface oxidation as a source of
  surface degradation factor for {Bi$_2$Se$_3$}},}\ }\bibfield  {booktitle}
  {\emph {\bibinfo {booktitle} {ACS Nano}},\ }\href {\doibase
  10.1021/nn200556h} {\bibfield  {journal} {\bibinfo  {journal} {ACS Nano}\
  }\textbf {\bibinfo {volume} {5}},\ \bibinfo {pages} {4698--4703} (\bibinfo
  {year} {2011})}\BibitemShut {NoStop}%
\bibitem [{\citenamefont {Ngabonziza}\ \emph {et~al.}(2015)\citenamefont
  {Ngabonziza}, \citenamefont {Heimbuch}, \citenamefont {Klaassen},
  \citenamefont {Stehno}, \citenamefont {Snelder}, \citenamefont {Solmaz},
  \citenamefont {Ramankutty}, \citenamefont {van Heumen}, \citenamefont
  {Koster},\ and\ \citenamefont {Golden}}]{Ngabonziza2015}%
  \BibitemOpen
  \bibfield  {author} {\bibinfo {author} {\bibfnamefont {P.}~\bibnamefont
  {Ngabonziza}}, \bibinfo {author} {\bibfnamefont {R.}~\bibnamefont
  {Heimbuch}}, \bibinfo {author} {\bibfnamefont {R.}~\bibnamefont {Klaassen}},
  \bibinfo {author} {\bibfnamefont {M.}~\bibnamefont {Stehno}}, \bibinfo
  {author} {\bibfnamefont {M.}~\bibnamefont {Snelder}}, \bibinfo {author}
  {\bibfnamefont {A.}~\bibnamefont {Solmaz}}, \bibinfo {author} {\bibfnamefont
  {S.}~\bibnamefont {Ramankutty}}, \bibinfo {author} {\bibfnamefont
  {E.}~\bibnamefont {van Heumen}}, \bibinfo {author} {\bibfnamefont
  {G.}~\bibnamefont {Koster}}, \ and\ \bibinfo {author} {\bibfnamefont
  {M.}~\bibnamefont {Golden}},\ }\bibfield  {title} {\enquote {\bibinfo {title}
  {In-situ spectroscopy of intrinsic {Bi$_2$Te$_3$} topological insulator thin
  films and impact of extrinsic defects},}\ }\href@noop {} {\bibfield
  {journal} {\bibinfo  {journal} {arXiv preprint arXiv:1502.01185}\ } (\bibinfo
  {year} {2015})}\BibitemShut {NoStop}%
\bibitem [{\citenamefont {Harrison}\ \emph {et~al.}(2014)\citenamefont
  {Harrison}, \citenamefont {Zhou}, \citenamefont {Huo}, \citenamefont {Pushp},
  \citenamefont {Kellock}, \citenamefont {Parkin}, \citenamefont {Harris},
  \citenamefont {Chen},\ and\ \citenamefont {Hesjedal}}]{Harrison2014}%
  \BibitemOpen
  \bibfield  {author} {\bibinfo {author} {\bibfnamefont {S.~E.}\ \bibnamefont
  {Harrison}}, \bibinfo {author} {\bibfnamefont {B.}~\bibnamefont {Zhou}},
  \bibinfo {author} {\bibfnamefont {Y.}~\bibnamefont {Huo}}, \bibinfo {author}
  {\bibfnamefont {A.}~\bibnamefont {Pushp}}, \bibinfo {author} {\bibfnamefont
  {A.~J.}\ \bibnamefont {Kellock}}, \bibinfo {author} {\bibfnamefont
  {S.~S.~P.}\ \bibnamefont {Parkin}}, \bibinfo {author} {\bibfnamefont {J.~S.}\
  \bibnamefont {Harris}}, \bibinfo {author} {\bibfnamefont {Y.}~\bibnamefont
  {Chen}}, \ and\ \bibinfo {author} {\bibfnamefont {T.}~\bibnamefont
  {Hesjedal}},\ }\bibfield  {title} {\enquote {\bibinfo {title} {Preparation of
  layered thin film samples for angle-resolved photoemission spectroscopy},}\
  }\href {\doibase http://dx.doi.org/10.1063/1.4896632} {\bibfield  {journal}
  {\bibinfo  {journal} {Applied Physics Letters}\ }\textbf {\bibinfo {volume}
  {105}},\ \bibinfo {pages} {121608} (\bibinfo {year} {2014})}\BibitemShut
  {NoStop}%
\bibitem [{\citenamefont {Virwani}\ \emph {et~al.}(2014)\citenamefont
  {Virwani}, \citenamefont {Harrison}, \citenamefont {Pushp}, \citenamefont
  {Topuria}, \citenamefont {Delenia}, \citenamefont {Rice}, \citenamefont
  {Kellock}, \citenamefont {Collins-McIntyre}, \citenamefont {Harris},
  \citenamefont {Hesjedal},\ and\ \citenamefont {Parkin}}]{Virwani2014}%
  \BibitemOpen
  \bibfield  {author} {\bibinfo {author} {\bibfnamefont {K.}~\bibnamefont
  {Virwani}}, \bibinfo {author} {\bibfnamefont {S.~E.}\ \bibnamefont
  {Harrison}}, \bibinfo {author} {\bibfnamefont {A.}~\bibnamefont {Pushp}},
  \bibinfo {author} {\bibfnamefont {T.}~\bibnamefont {Topuria}}, \bibinfo
  {author} {\bibfnamefont {E.}~\bibnamefont {Delenia}}, \bibinfo {author}
  {\bibfnamefont {P.}~\bibnamefont {Rice}}, \bibinfo {author} {\bibfnamefont
  {A.}~\bibnamefont {Kellock}}, \bibinfo {author} {\bibfnamefont
  {L.}~\bibnamefont {Collins-McIntyre}}, \bibinfo {author} {\bibfnamefont
  {J.}~\bibnamefont {Harris}}, \bibinfo {author} {\bibfnamefont
  {T.}~\bibnamefont {Hesjedal}}, \ and\ \bibinfo {author} {\bibfnamefont
  {S.}~\bibnamefont {Parkin}},\ }\bibfield  {title} {\enquote {\bibinfo {title}
  {Controlled removal of amorphous {Se} capping layer from a topological
  insulator},}\ }\href@noop {} {\bibfield  {journal} {\bibinfo  {journal}
  {Applied Physics Letters}\ }\textbf {\bibinfo {volume} {105}},\ \bibinfo
  {pages} {241605} (\bibinfo {year} {2014})}\BibitemShut {NoStop}%
\bibitem [{\citenamefont {Huefner}(2003)}]{Huefner2003}%
  \BibitemOpen
  \bibfield  {author} {\bibinfo {author} {\bibfnamefont {S.}~\bibnamefont
  {Huefner}},\ }\href {\doibase 10.1007/978-3-662-09280-4} {\emph {\bibinfo
  {title} {Photoelectron Spectroscopy: Principles and Applications}}},\
  \bibinfo {edition} {3rd}\ ed.,\ Advanced Texts in Physics\ (\bibinfo
  {publisher} {Springer-Verlag Berlin Heidelberg},\ \bibinfo {year} {2003})\
  p.\ \bibinfo {pages} {662}\BibitemShut {NoStop}%
\bibitem [{\citenamefont {Moulder}\ and\ \citenamefont
  {Chastain}(1992)}]{Moulder1992}%
  \BibitemOpen
  \bibfield  {author} {\bibinfo {author} {\bibfnamefont {J.}~\bibnamefont
  {Moulder}}\ and\ \bibinfo {author} {\bibfnamefont {J.}~\bibnamefont
  {Chastain}},\ }\href {https://books.google.de/books?id=A_XGQgAACAAJ} {\emph
  {\bibinfo {title} {Handbook of X-ray Photoelectron Spectroscopy: A Reference
  Book of Standard Spectra for Identification and Interpretation of XPS
  Data}}}\ (\bibinfo  {publisher} {Physical Electronics Division, Perkin-Elmer
  Corporation},\ \bibinfo {year} {1992})\BibitemShut {NoStop}%
\bibitem [{\citenamefont {Park}\ \emph {et~al.}(2015)\citenamefont {Park},
  \citenamefont {Soh}, \citenamefont {Aeppli}, \citenamefont {Feng},
  \citenamefont {Ou}, \citenamefont {He},\ and\ \citenamefont
  {Xue}}]{Park2015}%
  \BibitemOpen
  \bibfield  {author} {\bibinfo {author} {\bibfnamefont {J.}~\bibnamefont
  {Park}}, \bibinfo {author} {\bibfnamefont {Y.-A.}\ \bibnamefont {Soh}},
  \bibinfo {author} {\bibfnamefont {G.}~\bibnamefont {Aeppli}}, \bibinfo
  {author} {\bibfnamefont {X.}~\bibnamefont {Feng}}, \bibinfo {author}
  {\bibfnamefont {Y.}~\bibnamefont {Ou}}, \bibinfo {author} {\bibfnamefont
  {K.}~\bibnamefont {He}}, \ and\ \bibinfo {author} {\bibfnamefont {Q.-K.}\
  \bibnamefont {Xue}},\ }\bibfield  {title} {\enquote {\bibinfo {title}
  {Crystallinity of tellurium capping and epitaxy of ferromagnetic topological
  insulator films on {SrTiO$_3$}},}\ }\href
  {http://dx.doi.org/10.1038/srep11595} {\bibfield  {journal} {\bibinfo
  {journal} {Sci. Rep.}\ }\textbf {\bibinfo {volume} {5}} (\bibinfo {year}
  {2015})}\BibitemShut {NoStop}%
\bibitem [{\citenamefont {Tjeng}\ \emph {et~al.}(1997)\citenamefont {Tjeng},
  \citenamefont {Hesper}, \citenamefont {Heessels}, \citenamefont {Heeres},
  \citenamefont {Jonkman},\ and\ \citenamefont {Sawatzky}}]{Tjeng1997}%
  \BibitemOpen
  \bibfield  {author} {\bibinfo {author} {\bibfnamefont {L.}~\bibnamefont
  {Tjeng}}, \bibinfo {author} {\bibfnamefont {R.}~\bibnamefont {Hesper}},
  \bibinfo {author} {\bibfnamefont {A.}~\bibnamefont {Heessels}}, \bibinfo
  {author} {\bibfnamefont {A.}~\bibnamefont {Heeres}}, \bibinfo {author}
  {\bibfnamefont {H.}~\bibnamefont {Jonkman}}, \ and\ \bibinfo {author}
  {\bibfnamefont {G.}~\bibnamefont {Sawatzky}},\ }\bibfield  {title} {\enquote
  {\bibinfo {title} {Development of the electronic structure in a {K}-doped
  {C60} monolayer on a {Ag}(1 1 1) surface},}\ }\href
  {http://www.sciencedirect.com/science/article/pii/S0038109897001269}
  {\bibfield  {journal} {\bibinfo  {journal} {Solid State Communications}\
  }\textbf {\bibinfo {volume} {103}},\ \bibinfo {pages} {31--35} (\bibinfo
  {year} {1997})}\BibitemShut {NoStop}%
\bibitem [{\citenamefont {Hesper}, \citenamefont {Tjeng},\ and\ \citenamefont
  {Sawatzky}(1997)}]{Hesper1997}%
  \BibitemOpen
  \bibfield  {author} {\bibinfo {author} {\bibfnamefont {R.}~\bibnamefont
  {Hesper}}, \bibinfo {author} {\bibfnamefont {L.~H.}\ \bibnamefont {Tjeng}}, \
  and\ \bibinfo {author} {\bibfnamefont {G.~A.}\ \bibnamefont {Sawatzky}},\
  }\bibfield  {title} {\enquote {\bibinfo {title} {Strongly reduced band gap in
  a correlated insulator in close proximity to a metal},}\ }\href
  {http://stacks.iop.org/0295-5075/40/i=2/a=177} {\bibfield  {journal}
  {\bibinfo  {journal} {EPL (Europhysics Letters)}\ }\textbf {\bibinfo {volume}
  {40}},\ \bibinfo {pages} {177--} (\bibinfo {year} {1997})}\BibitemShut
  {NoStop}%
\bibitem [{\citenamefont {Musket}(1978)}]{Musket1978}%
  \BibitemOpen
  \bibfield  {author} {\bibinfo {author} {\bibfnamefont {R.}~\bibnamefont
  {Musket}},\ }\bibfield  {title} {\enquote {\bibinfo {title} {Studies of clean
  and oxidized tellurium surfaces},}\ }\href
  {http://www.sciencedirect.com/science/article/pii/0039602878900377}
  {\bibfield  {journal} {\bibinfo  {journal} {Surface Science}\ }\textbf
  {\bibinfo {volume} {74}},\ \bibinfo {pages} {423--435} (\bibinfo {year}
  {1978})}\BibitemShut {NoStop}%
\bibitem [{\citenamefont {Lee}\ and\ \citenamefont {Geiss}(1983)}]{Lee1983}%
  \BibitemOpen
  \bibfield  {author} {\bibinfo {author} {\bibfnamefont {W.-Y.}\ \bibnamefont
  {Lee}}\ and\ \bibinfo {author} {\bibfnamefont {R.~H.}\ \bibnamefont
  {Geiss}},\ }\bibfield  {title} {\enquote {\bibinfo {title} {Degradation of
  thin tellurium films},}\ }\href {\doibase http://dx.doi.org/10.1063/1.332156}
  {\bibfield  {journal} {\bibinfo  {journal} {Journal of Applied Physics}\
  }\textbf {\bibinfo {volume} {54}},\ \bibinfo {pages} {1351--1357} (\bibinfo
  {year} {1983})}\BibitemShut {NoStop}%
\end{thebibliography}%

\onecolumngrid
\newpage


\pagenumbering{roman}
\renewcommand{\figurename}{FIG.\,\,S$\!\!$}
\makeatother
\setcounter{figure}{0}
\setcounter{page}{1}

\begin{flushleft}
\Large{\textsf{\textbf{Supplemental Material\\\medskip
 Protective  capping of  topological surface states of intrinsically insulating Bi$_2$Te$_3$}}} \\\bigskip
\indent \normalsize{\textsf{Katharina Hoefer, Christoph Becker, Steffen Wirth, and Liu Hao Tjeng\\
\textit{Max Planck Institute for
Chemical Physics of Solids, N\"{o}thnitzer Strasse 40, Dresden 01187, Germany}}}
\end{flushleft}

\begin{figure*}[htb]
\begin{center}
\centerline{\includegraphics[width=1\textwidth]{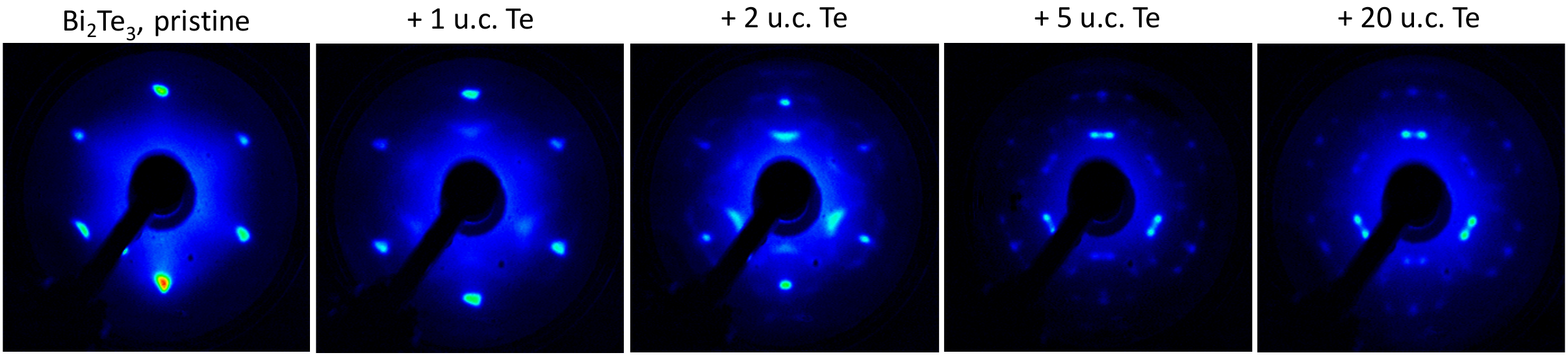}}
  \caption{LEED pattern recorded at 26\,eV electron energy of a pristine 20\,QL Bi$_2$Te$_3$ film   and the same film at different stages (1\,u.c. $\sim$6\,{\AA}, 2\,u.c., 5\,u.c. and 20\,u.c.)    of  Te capping.}\label{suppl_LEED}
\end{center}
\end{figure*}

\begin{figure*}[htb]
\begin{center}
\centerline{\includegraphics[width=0.55\textwidth]{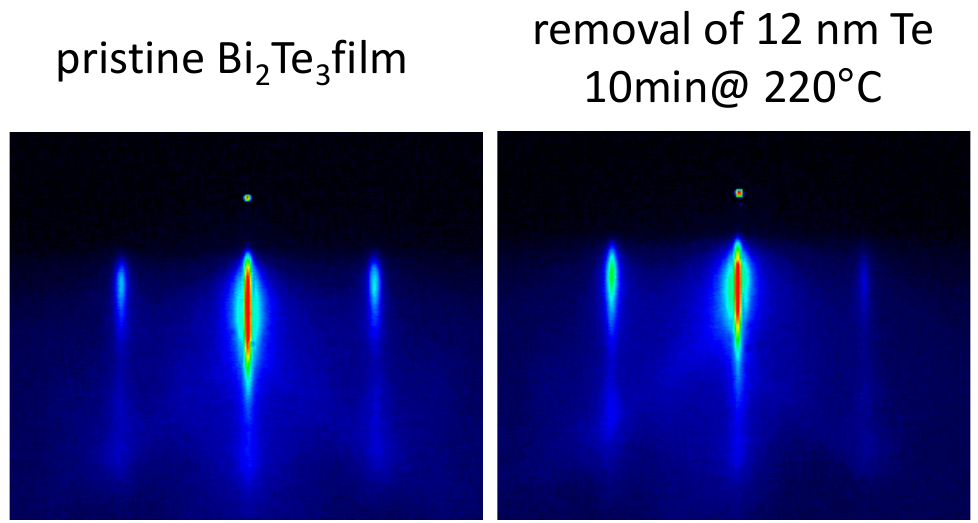}}
  \caption{RHEED pattern of a 20\,QL pristine Bi$_2$Te$_3$ film (left) and the same film after thermal desorption of 12\,nm Te capping for 10\,min at 220$^{\circ}$C.}\label{suppl_RHEED}
  \end{center}
\end{figure*}

\begin{figure*}[htb]
\begin{center}
\centerline{\includegraphics[width=0.65\textwidth]{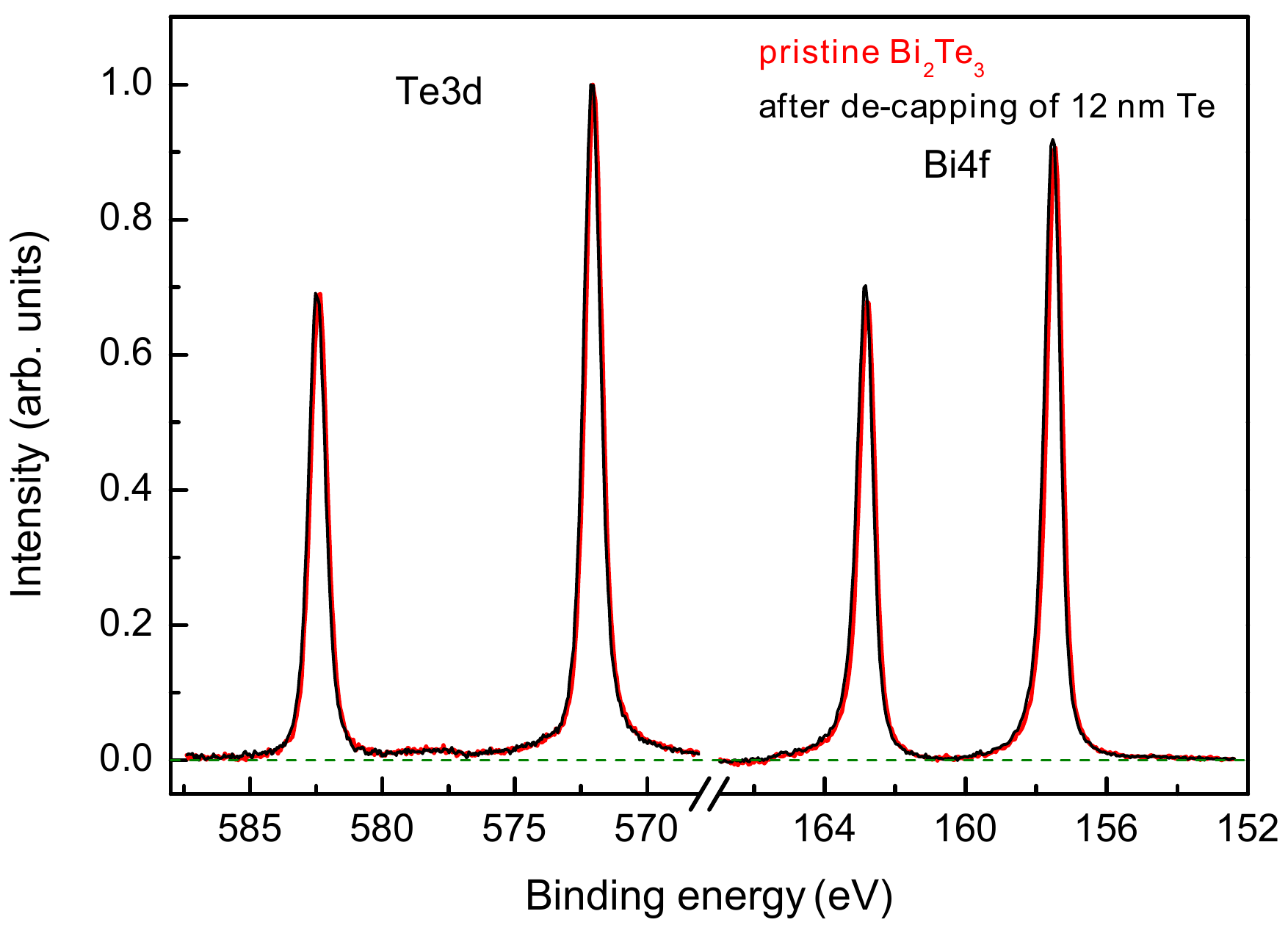}}
  \caption{XPS spectrum of a 20\,QL Bi$_2$Te$_3$ film, (a) pristine (red curve) and
                           (b)  after thermal desorption of  12\,nm Te capping.}\label{suppl_XPS}
  \end{center}
\end{figure*}

\end{document}